\documentclass[12pt]{iopart}
\usepackage[dvips]{graphicx}
\begin{document}
\jl{3}
\title[Hyperfine interactions in Fe-N and Fe-C fcc alloys]{Influence of carbon
and nitrogen on electronic structure and hyperfine interactions in
fcc iron-based alloys}

\author{A N Timoshevskii \dag\footnote[3]{To whom correspondence should be
addressed.}, V A Timoshevskii \ddag\ and B Z Yanchitsky\dag}

\address{\dag\ Institute of Magnetism, 36-b Vernadskii St., 252680 Kiev, Ukraine}

\address{\ddag\ Rostov State University of Transport, Narodnogo Opolcheniya 2,
Rostov-on-Don, 344038 Russia}

\begin{abstract}
Carbon and nitrogen austenites, modeled by Fe$_8$N and Fe$_8$C
superstructures are studied by full-potential LAPW method.
Structure parameters, electronic and magnetic properties as well
as hyperfine interaction parameters are obtained. Calculations
prove that Fe-C austenite can be successfully modeled by ordered
Fe$_8$C superstructure. The results show that chemical Fe-C bond
in Fe$_8$C has higher covalent part than in Fe$_8$N. Detailed
analysis of electric field gradient formation for both systems is
performed. The calculation of electric field gradient allow us to
carry out a good interpretation of M\"ossbauer spectra for Fe-C
and Fe-N systems.
\end{abstract}


\section{Introduction}\label{sec1}
Face centered cubic (fcc) iron-based alloys are widely used for
developing of stainless austenitic steels especially for using in
critical temperature ranges, aggressive environment and other
severe external conditions. Doping of these steels with light
interstitial impurities influence mechanics and kinetics of
structure phase transitions in Fe-based alloys. Nitrogen doping
enables to solve the problem of the strengthening of stainless
steels.

Carbon and nitrogen doping differently influence structure and
mechanical properties of austenites. Studying of this effect by
means of M\"ossbauer spectroscopy determined substantial
differences in spectra of carbon and nitrogen austenitic steels
(Sozinov \etal 1997, 1999). A lot of information, containing in
M\"ossbauer spectra as well as high accuracy of measurements lead
to the fact that detailed interpretation of these experimental
results appears to be a very complicated problem and needs
application of high-accuracy theoretical methods.

The role of \textit{ab initio} calculations in the area of solid
state research rapidly increased last years. Step by step
semiempirical methods, containing a lot of parameters, are
replaced by calculations from first principles. We observe the
extension of application field of \textit{ab initio} quantum
mechanical methods of calculations of electronic structure and
physical properties of solids. Last years these methods are more
often used for solving not only fundamental, but also applied
problems.

 In this paper we study the influence of carbon
and nitrogen on atomic and electronic structure of fcc-iron.
M\"ossbauer spectroscopy gives the most interesting data about
impurity distributions, electronic structure and magnetic
interactions in solids. The study of shifts and splitting of
nuclear energy levels gives information about symmetry of charge
distributions near the nucleus, electronic configurations of atoms
and ions as well as about peculiarities of atomic structure of
solids.  We believe that detailed interpretation of these
experimental data using single up-to-date \textit{ab-initio}
approach is an important step in investigation of the influence
of light impurities (C,N) in case of real Fe(fcc) - based alloys.

\section{Atomic structure} \label{sec2}
Up to now there is no full understanding of influence of carbon
and nitrogen atoms on atomic and electronic structure of fcc Fe-C
and Fe-N alloys. Based on M\"ossbauer spectroscopy data different
authors made different conclusions about atomic structure of
carbon and nitrogen Fe-based fcc alloys. For example, the group
of Genin concluded that carbon distribution in fcc Fe-C alloy is
close to the ordered Fe$_8$C$_{1-x}$ structure (Bauer \etal
1990). On the other hand, the group of Gavriljuk using another
method of M\"ossbauer spectrum deconvolution and after performing
Monte Carlo computer simulations showed that formation of the
ordered Fe$_8$C$_{1-x}$ structure is improbable (Sozinov \etal
1997).

In this paper we are trying to solve two problems: to study
differences in electronic structure of carbon and nitrogen
austenites, and to perform detailed interpretation of M\"ossbauer
experimental data for Fe-N and Fe-C alloys. As an object of our
investigations we chose model Fe$_8$C and Fe$_8$N fcc-type
superstructures. The same type of C and N ordering in this
structures allow us to determine peculiarities of electronic
structure and hyperfine interaction parameters, connected only
with different influence of two types of impurity atoms.

Calculations were performed by applying highly accurate WIEN97
programme package (Blaha \etal 1999),  which employs
full-potential LAPW method (Singh 1994a). Generalized gradient
approximation (GGA) according to Perdew-Burke-Ernzerhof (Perdew
\etal 1996) model was used for exchange-correlation potential. The
value of plane-wave cutoff was $R_{mt}\times K_{max}=8.4$ which
corresponds to about 190 plane waves per atom in the basis set.
Inside atomic spheres the wave function was decomposed up to
$l_{max}=12$. Charge density and potential was decomposed inside
atomic spheres using lattice harmonics basis up to $L_{max}=6$. In
the interstitial region Fourier expansion was used with 847
coefficients. Calculations were performed for 3000 K-points in
the Brrillouin zone. The radii of atomic spheres were chosen as
1.9 a.u. for Fe atoms and 1.6 a.u for N(C) atoms. Standard LAPW
basis set was expanded by including local orbitals (Singh 1994b)
for Fe$3s3p$ - states and N(C)$2s$ - states. The values of all
parameters were tested for convergence and ensure accuracy of 0.1
mRy in total energy of the system. In order to check the role of
local magnetic moments on Fe atoms, all calculations were
performed using spin-polarized approximation.

\begin{figure}[!tb] \label{cell}
\begin{center}
\includegraphics[scale=0.5,angle=0]{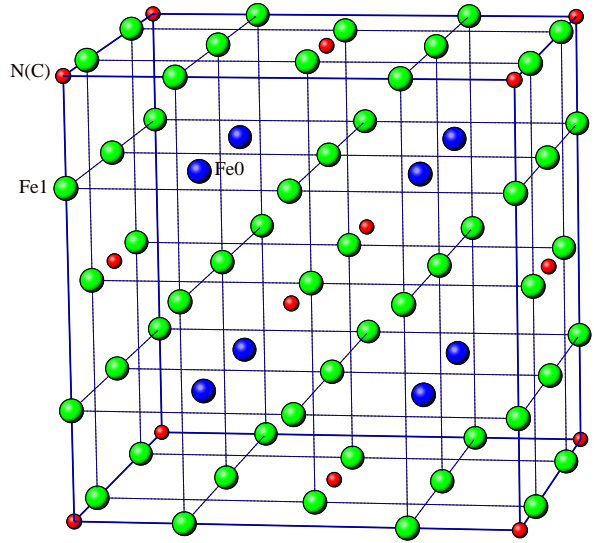}
\end{center}
\caption{Ordered Fe$_8$A superstructures (A=C,N). Atoms Fe$_1$
have one impurity atom in the first coordination sphere. Atoms
Fe$_0$ have no impurity in the nearest neighbourhood.}
\end{figure}

Figure 1 presents Fe$_8$A supercell (Fm$\overline{3}$m, No.225),
which was used for calculations. The structure has two symmetry
types of Fe atoms: Fe$_1$ type forms octahedron around impurity
atom (has one impurity atom in the first coordination sphere);
Fe$_0$ type has no impurity atoms in the first coordination
sphere. Before we performed the final electronic structure and
hyperfine interaction parameters calculations, total energy
minimization procedure was performed to obtain equilibrium
positions of atoms. The lattice constant as well as the size of
Fe$_1$-octahedron were varied to achieve total energy minimum.
Optimization results were approximated by second-order
polynomials using least square fit method.  Then we analytically
found the values of lattice parameters and positions of
Fe-octahedron atoms. Calculated and experimental values of
lattice parameters as well as calculated distance between Fe and
C(N) atoms are presented in \tref{tab1}. Optimization results for
iron nitride Fe$_4$N are also presented. Good agreement with
experiment shows that the method can be successfully applied for
iron-nitrogen (iron-carbon) compounds.

\section{Electronic structure}\label{sec3}

Figure 2 presents total and partial density of states for Fe, C
and N atoms in Fe$_8$A compounds. The main difference in
electronic structure of two compounds is the energy position of
impurity $2p$ states. In Fe$_8$C carbon $2p$ states are located
right below the $d$-band of iron. In Fe$_8$N nitrogen $2p$ states
are located 2 eV below the bottom of Fe $d$-band. This difference
in energy positions of C and N $2p$ states leads to essential
difference in the character of chemical bonding in these two
compounds. In Fe$_8$C we see stronger $p-d$ hybridization compare
to Fe$_8$N compound. The width of N $2p$-band is twice shorter
than corresponding $2p$-band of carbon. This probably leads to
much higher Fe-C interaction compare to Fe-N interaction.

\begin{table}
\caption{Structure and magnetic parameters for Fe$_8$N, Fe$_8$C
and Fe$_4$N compounds. Fe-N(C) is the distance between Fe and
impurity atom (a.u.) and $\mu$ is local magnetic moment
($\mu_B$).}\label{tab1}
\begin{indented}
\lineup
\item[]\begin{tabular}{@{}*{8}{l}}
\br &Lattice&Fe-A&&&&&$\mu$(aver./\\
&parameter&distance&$\mu_{Fe_0}$&$\mu_{Fe_1}$&$\mu_{Fe_2}$&
$\mu_{N,C}$&\0\0atom)\\ \mr

Fe$_8$N (theor.)&13.946&3.54&2.78&2.09&---&$-0.11$&2.25\\
\0\0\0\0\0 (expt)&13.849$^{\rm a}$&&&&&&\\

Fe$_8$C (theor.)&13.981&3.58&2.71&2.12&---&$-0.16$&2.25\\
\0\0\0\0\0 (expt)&13.839$^{\rm a}$&&&&&&\\

Fe$_4$N (theor.)&\07.164&3.582&2.90&---&2.31&$\m 0.04$&2.50\\
\0\0\0\0\0 (expt)&\07.171$^{\rm b}$&3.585$^{\rm b}$&3.00$^{\rm
c}$&---&2.00$^{\rm c}$&&2.21$^{\rm c}$\\ \br
\end{tabular}
\item[] $^{\rm a}$ Cheng \etal (1990).
\item[] $^{\rm b}$ Jacobs \etal (1995).
\item[] $^{\rm c}$ Frazer (1958).
\end{indented}
\end{table}

\begin{figure}[!tb] \label{dos}
\begin{center}
\includegraphics[scale=0.45,angle=0]{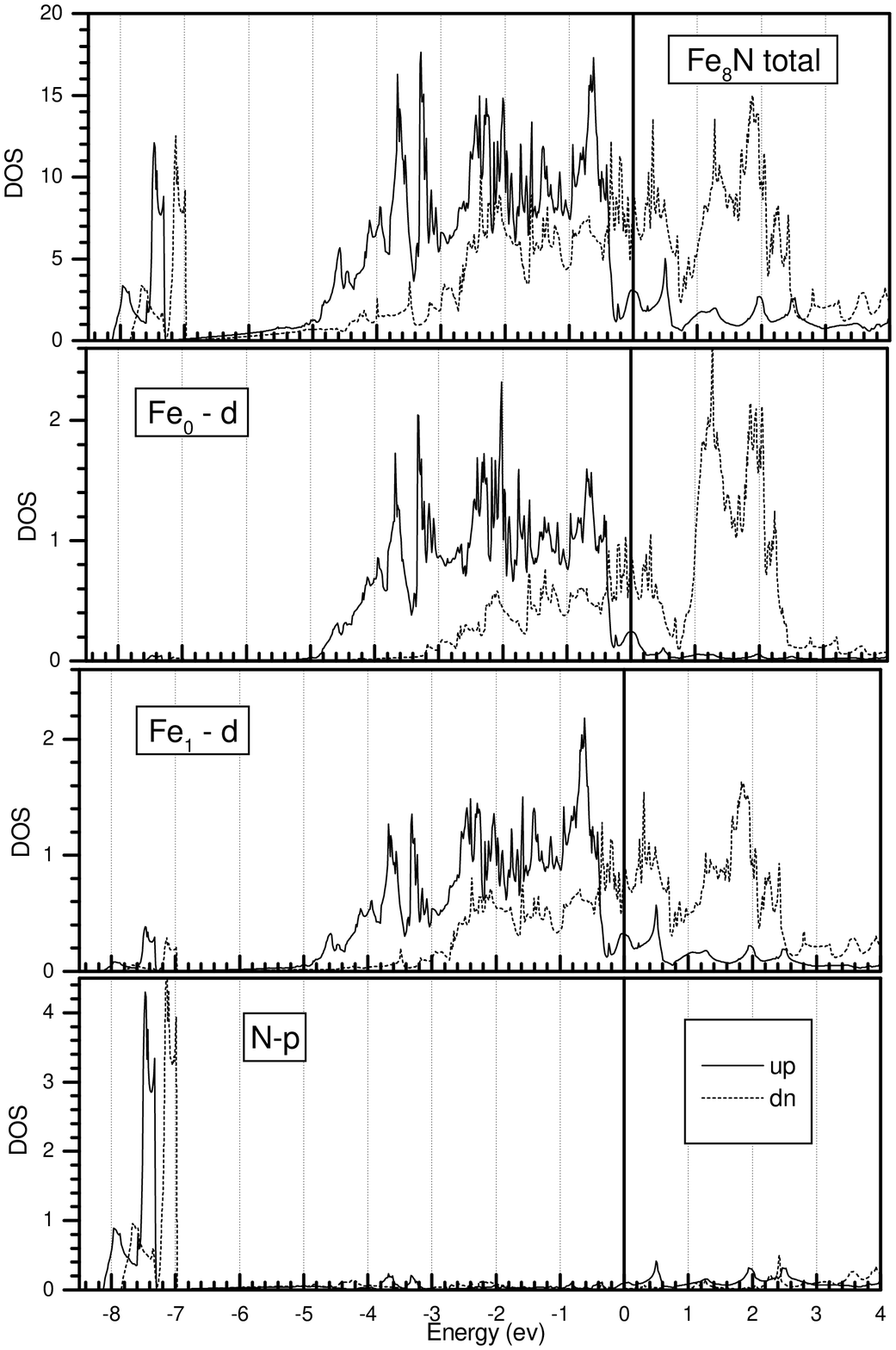}
\includegraphics[scale=0.45,angle=0]{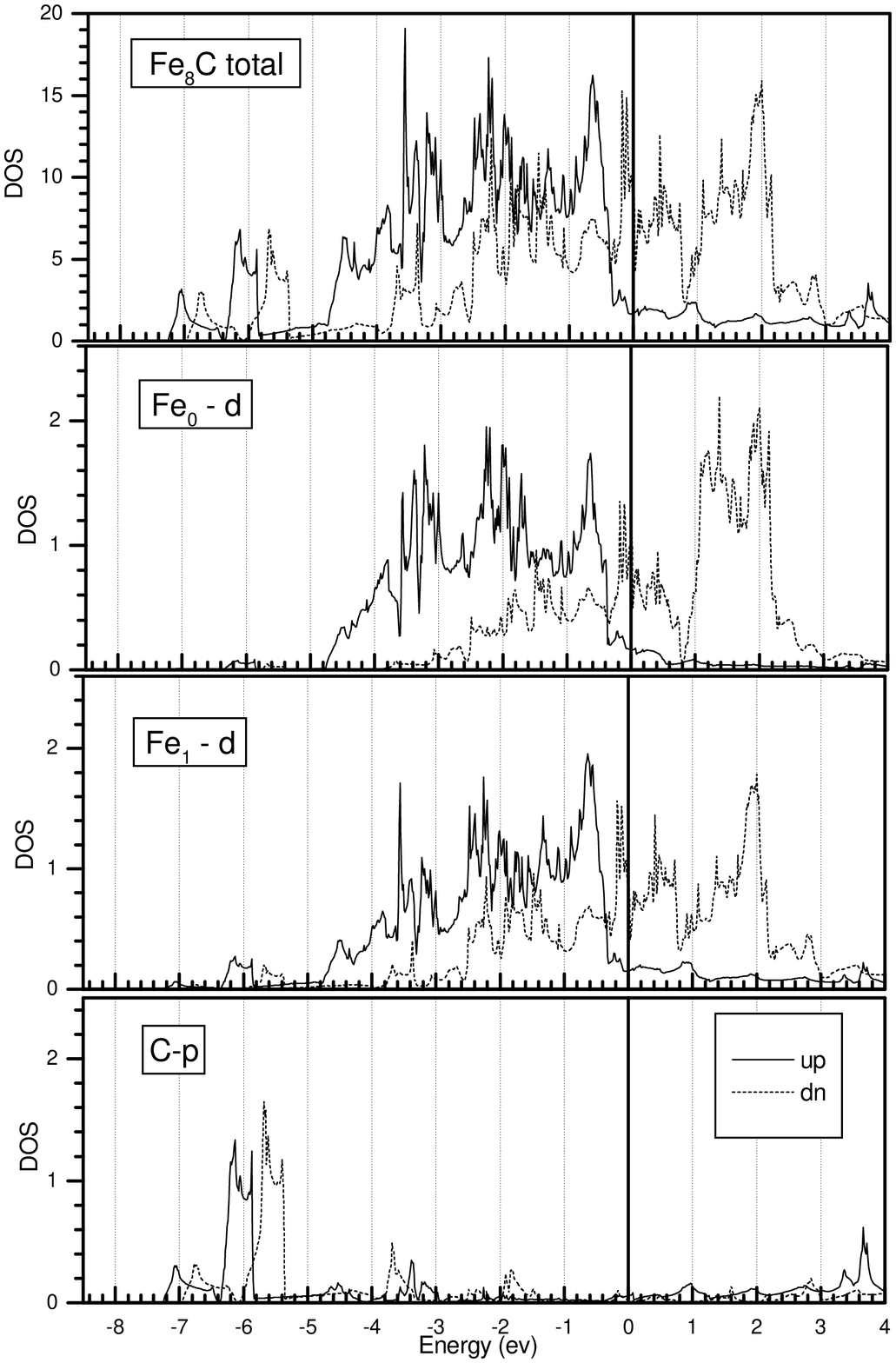}
\end{center}
\caption{Density of states (DOS) for Fe$_{8}$A (A=C,N) compounds}
\end{figure}

The difference in energy distribution of valent electrons leads to
considerable difference in electronic density distribution in
Fe$_8$N and Fe$_8$C. Figure 3 presents electronic density
distribution in (100) plane for two energy intervals: $2p$ states
of impurity atoms and $3d$ states of iron in Fe$_8$N and Fe$_8$C
compounds. Nitrogen $2p$ electrons are much more localized
(figure 3a) compare to $2p$ electrons of carbon (figure 3c).
Charge distribution around Fe$_1$ atom in Fe$_8$C is more
symmetric (figure 3d) compare to corresponding charge
distribution in Fe$_8$N (figure 3b). It should be noted that
impurity atoms cause asymmetric charge distribution around Fe$_1$
atom. This also leads to asymmetric charge distribution of $d$
electrons around Fe$_1$ atom. These effects greatly influence
electric field gradient (EFG) at Fe$_1$ nuclei, which can be
measured as quadrupole splitting in M\"ossbauer spectra of these
two compounds.
\begin{figure}[!tb] \label{eplot}
\begin{center}
\includegraphics[scale=0.5,angle=0]{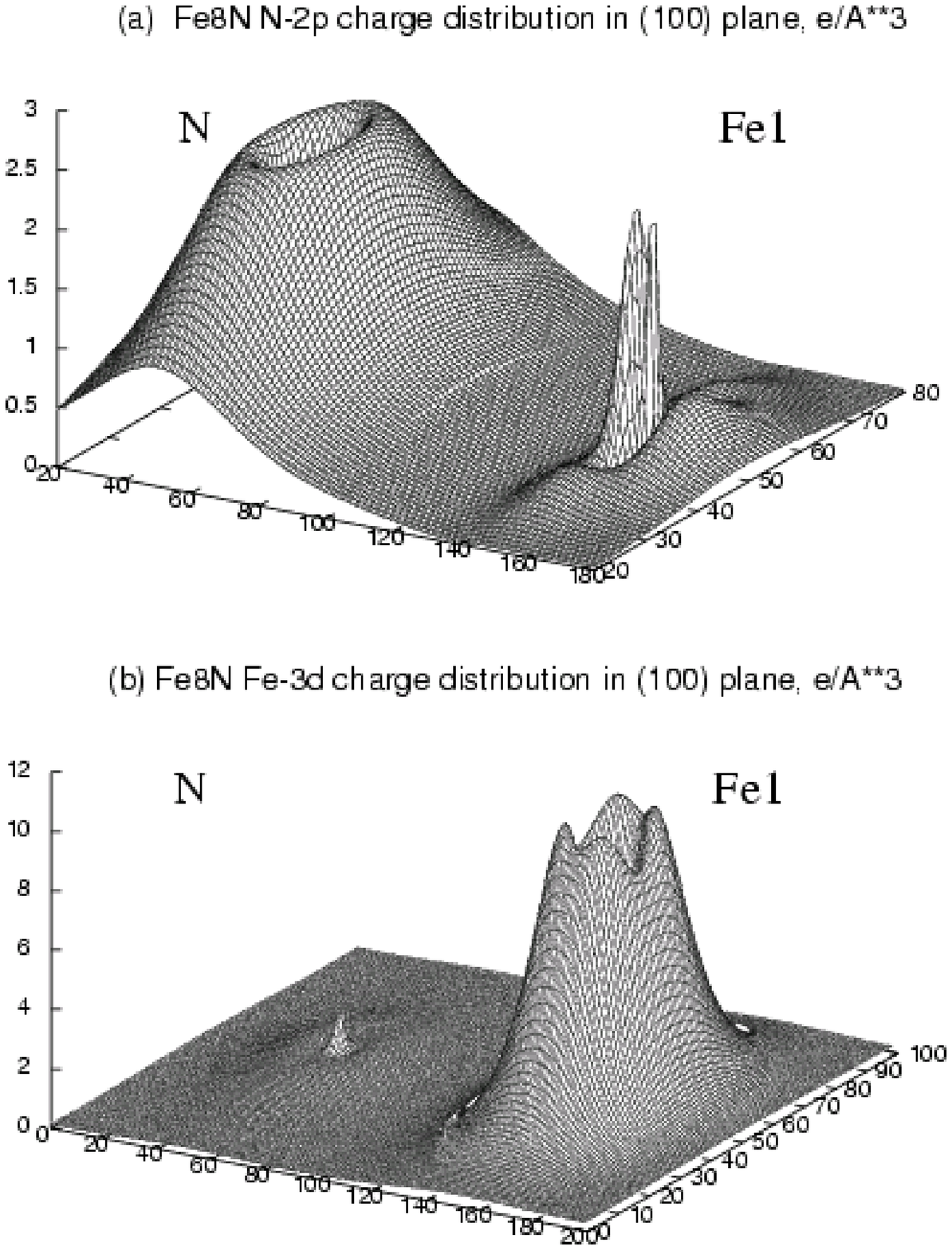}
\includegraphics[scale=0.5,angle=0]{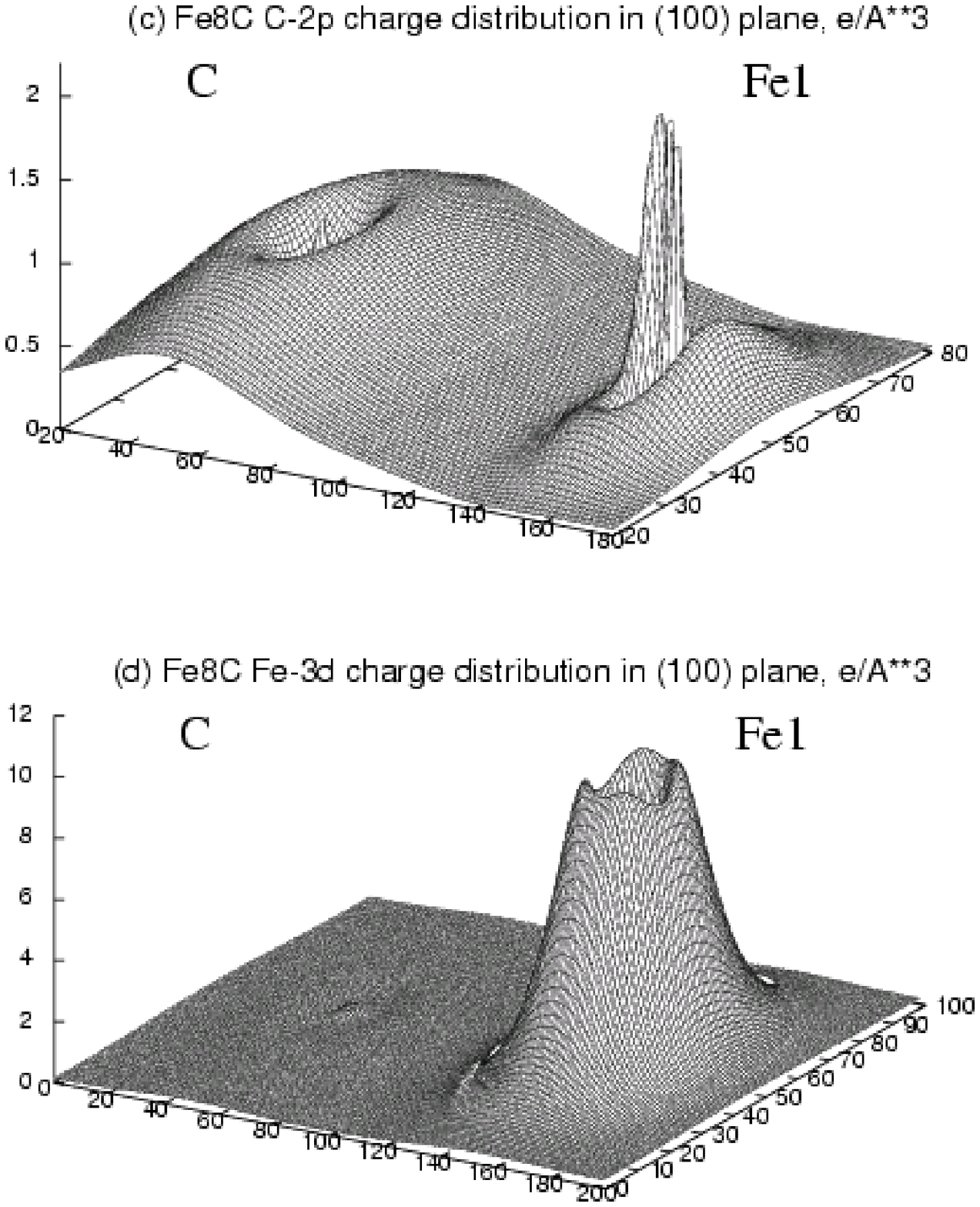}
\end{center}
\caption{Fe$_8$N and Fe$_8$C charge distribution in (100) plane
($e/\AA^3$) for different energy regions.}
\end{figure}

\section{Hyperfine interactions in Fe$_8$N and Fe$_8$C}
\label{sec4}

For our structures $^{57}$Fe nucleus quadrupole splitting is given
by the following expression:
\begin{equation}\label{qs1}
\Delta=\frac{1}{2}eQV_{zz},
\end{equation}
where $Q$ - quadrupole moment of the nucleus and $V_{zz}$ -
principal component of electric field gradient (EFG) tensor.

In our calculations we used Q$^{57}$Fe =0.16$b$, which was
determined by Dufek, Blaha and Schwarz (1995) by comparing
experimental quadrupole splitting and calculated EFG values for a
large number of different Fe compounds. EFG is calculated on
\textit{ab-initio} basis directly from electronic density
distribution using method, developed by Blaha, Schwarz and Herzig
(1985).

\Tref{tab2} presents our theoretical results for quadrupole
splitting as well as experimental data for carbon and nitrogen
austenites, obtained by several groups. We see that reasonable
agreement with experiment is obtained for Fe-C alloy. This leads
to conclusion that local environment of Fe$_1$ atom in real alloy
is close to Fe$_1$ environment in our model Fe$_8$C
superstructure. This is also supported by the fact that according
to experimental data (Bauer \etal 1988, Gavriljuk and Nadutov
1983, Oda \etal 1994) Fe-C austenite has only Fe$_0$ and Fe$_1$
atoms, which corresponds to Fe configurations in our model
superstructure. On the other side, in Fe-N austenite M\"ossbauer
spectrum is formed by contributions from Fe$_0$, Fe$_1$ and
Fe$_{2-180^o}$ (dumb-bell configuration) atoms (Oda \etal 1990,
Foct \etal 1987, Gavriljuk \etal 1990). Both absence of
Fe$_{2-180^o}$ atoms in our structure and bad agreement with
experiment shows that Fe-N austenite can not be modeled by Fe$_8$N
superstructure.

Our calculation shows that contribution to V$_{zz}$ (the principal
component of EFG tensor) from the regions outside the spheres
(lattice EFG) is 5\%, 3\% and 10\% for Fe$_8$N, Fe$_8$C and
Fe$_4$N respectively. So, for understanding of the origin of the
EFG we focus on its main component, the valence EFG, which
originates from the space inside the atomic spheres. The
ingredients for the calculation of the valence EFG are the density
coefficients $\rho_{2M}$, which originate from two radial wave
functions with $l$ and $l'$:
\begin{equation}\label{qs2}
\rho_{LM}(r)=\sum_{E<E_F}\sum_{l,m}\sum_{l',m'}R_{lm}(r)R_{l'm'}(r)G_{Lll'}^{Mmm'},
\end{equation}
where $R_{lm}(r)$ - LAPW radial wave functions, $G_{Lll'}^{Mmm'}$
- Gaunt integrals.

We performed comparative analysis of EFG formation for Fe$_8$N and
Fe$_8$C structures. \Tref{tab2} and \tref{tab3} show that total
EFG and quadrupole splitting is almost 3 times larger in Fe$_8$C
structure compare to Fe$_8$N. By performing the analysis of
mechanisms of EFG formation, we tried to find out the reasons of
this considerable difference. First of all it is obvious that
contribution from Fe $3s3p$ energy interval remains constant while
going from nitrogen to carbon compound. This means that
redistribution of electronic density in this region is not
considerable and can not influence total EFG increasing. We
observe increase of negative contribution from C $2s$ interval,
but it is also not considerable. Moreover, the sum of
contributions from impurity $2s2p$ energy interval remains almost
constant while going from nitrogen to carbon. This means that the
main reason of EFG increasing is redistribution of charge density
in Fe $3d$ band (and partially in impurity $2p$ band).

\begin{table}
\caption{Calculated and experimental values of quadrupole
splitting $\Delta$(mm/s) in M\"ossbauer spectra for Fe-N, Fe-C fcc
alloys.}\label{tab2}
\begin{indented}
\lineup
\item[]\begin{tabular}{@{}*{8}{l}}
\br &Fe$_4$N&Fe$_8$N&Fe$_{10.1}$N&Fe$_{11}$N&Fe$_8$C&Fe$_{11.3}$C
&Fe$_{11.5}$C\\ \mr

$\Delta$ (theor.)&0.50&0.18&&&0.53&\\

\0\0 (expt)&0.50$^{\rm a}$&&0.39$^{\rm b}$&0.25$^{\rm
c}$&&0.64$^{\rm e}$&0.67$^{\rm g}$\\

\0\0\0\0\0 &&&&0.39$^{\rm d}$&&0.63$^{\rm f}$\\ \br
\end{tabular}
\item[] $^{\rm a}$ Foct (1974).
\item[] $^{\rm b}$ Oda \etal (1990).
\item[] $^{\rm c}$ Foct \etal (1987).
\item[] $^{\rm d}$ Gavriljuk \etal (1990).
\item[] $^{\rm e}$ Genin and Flinn (1968).
\item[] $^{\rm f}$ DeCristofaro and Kaplow (1977).
\item[] $^{\rm g}$ Oda \etal (1994).
\end{indented}
\end{table}

In Fe$_8$N $2p3d$-band EFG contribution is positive, which
decreases total EFG. In Fe$_8$C this contribution is negative and
increases total EFG. In Fe$_8$C charge distributions in $2p$ and
$3d$ bands is more symmetric than in Fe$_8$N. While going from N
to C the relative increasing of symmetry of $3d$ charge
distribution is much larger than relative increasing of symmetry
of $2p$ charge distribution, which leads to smaller EFG
compensation and, as a result, to negative contribution from
$2p3d$-band.

Let us now consider partial EFG contributions ($p-p$, $d-d$)
inside Fe$_1$ sphere (other symmetry contributions like $s-d$,
etc. are negligibly small and can not determine EFG formation).
While going from N to C we observe relatively small increase of
the absolute value of $p-p$ contribution and quite large decrease
of $d-d$ contribution. As far as $d-d$ contribution is mostly
formed by own Fe$3d$ electrons, we can make a conclusion that
while going from N to C the increasing of space symmetry of Fe
$3d$ electrons causes stronger influence of the asymmetric
distribution of $2s$ and $2p$ electrons of the impurity, located
in Fe$_1$ sphere.

\begin{table}
\caption{EFG analysis for Fe$_1$ atom in Fe$_8$N and Fe$_8$C
superstructures. Contributions to valent EFG (10$^{21}$ V/m$^2$)
from different energy bands are presented.}\label{tab3}
\begin{indented}
\lineup
\item[]\begin{tabular}{@{}*{10}{l}}
\br Fe$_8$N&$V_{zz}^{p-p}$&$V_{zz}^{d-d}$&$V_{zz}^{val}$&\0\0\0&
Fe$_8$C&$V_{zz}^{p-p}$&$V_{zz}^{d-d}$&$V_{zz}^{val}$\\ \mr
Fe$3s3p$&$+2.448$&$\m0.000$&$+2.417$&\0\0\0&Fe$3s3p$&$+2.455$&$\m0.000$&$+2.434$\\
N$2s$&$-4.369$&$-0.174$&$-4.554$&\0\0\0&C$2s$&$-4.822$&$-0.354$&$-5.194$\\
N$2p$&$-4.552$&$-2.467$&$-7.067$&\0\0\0&C$2p$&$-2.479$&$-3.177$&$-5.695$\\
Fe$3d$&$+0.686$&$+7.310$&$+8.074$&\0\0\0&Fe$3d$&$-1.200$&$+6.309$&$+5.166$\\
\mr $Valent$&&&&&$Valent$&&\\

$band$&$-5.787$&$+4.669$&$-1.130$&&$band$&$-6.046$&$+2.778$&$-3.289$\\
\br
\end{tabular}
\end{indented}
\end{table}

Finally we tried to analyze the reasons of symmetry increase of
$3d$-band charge distribution in Fe$_8$C. Let us consider partial
EFG contributions only for $3d$-band. In Fe$_8$N $p-p$ and $d-d$
contributions are both positive and decrease total (negative) EFG.
In Fe$_8$C situation is different: $p-p$ contribution becomes
negative and its absolute value becomes twice larger. This
decreases positive $d-d$ contribution and causes the increase of
total EFG.

\section{Summary}
The results of our full-potential band structure calculations of
Fe$_8$N and Fe$_8$C structures lead to the following conclusions.
We observe that in Fe$_8$C chemical Fe-C bonding is more covalent
compare to Fe-N bonding in Fe$_8$N. Fe-C austenite can be
successfully modeled by ordered Fe$_8$C superstructure,
containing iron atoms in Fe$_0$ and Fe$_1$ configurations.
Calculated quadrupole splitting is in good agreement with
experimental values.

We performed detailed analysis of EFG formation on Fe nuclei in
this structure. The main EFG contribution in made by carbon 2s
and 2p states. The contributions of these two states appear to be
almost equal. Carbon states contributions are greatly canceled by
asymmetry of space distribution of Fe$3d$-electrons.

We also calculated structure, magnetic and hyperfine interaction
parameters for ordered iron nitride Fe${_4}$N structure. The
results of our calculations are in excellent agreement with
experimental data.

We come to conclusion that Fe-N austenite can not be modeled by
Fe$_8$N structure. The results of our calculations show that
nitrogen distribution in real Fe-N alloy is different from
nitrogen distribution in Fe$_8$N structure.

\References

\item[] Bauer Ph, Uwakweh O N C and Genin J M R 1988 {\it Hyperfine
Interactions} {\bf 41} 555
\item[] Bauer Ph, Uwakweh O N C and Genin J M R 1990 {\it Metall. Trans.}
{\bf A21} 589
\item[] Blaha P, Schwarz K and Luitz J 1999 {\it WIEN97, A Full
Potential Linearized Augmented Plane Wave Package for Calculating
Crystal Properties} (Techn. Universitat Wien, Austria ISBN
3-9501031-0-4)
\item[] Blaha P and Schwarz K 1983 {\it Int. J. Quantum
Chem.} {\bf XXIII} 1535
\item[] Blaha P, Schwarz K and Herzig P 1985 \PRL {\bf 54} 1192
\item[] Dufek P, Blaha P and Schwarz K 1995 \PRL {\bf 75} 3545
\item[] Foct J, Rochegude P and Hendry A 1988 \AM {\bf 36} 501
\item[] Gavriljuk V G and Nadutov V M 1983 {\it Fiz. Metall.
Metalloved.} {\bf 55} 520
\item[] Gavriljuk V G, Nadutov V M and Gladun O 1990 {\it Physics
Metals Metallogr} {\bf 3} 128
\item[] Oda K, Fujimura H and Ino H 1994 \JPCM {\bf 6} 679
\item[] Perdew J P, Burke S and Ernzerhof M 1996 \PRL {\bf 77} 3865
\item[] Singh D 1994a {\it Plane Waves, Pseudopotentials and the
LAPW method} (Kluwer Academic)
\item[] Singh D 1994b \PR {\it B} {\bf 43} 6388
\item[] Sozinov A L, Balanyuk A G and Gavriljuk V G 1997 {\it Acta
Mater.} {\bf 45} 225
\item[] Sozinov A L, Balanyuk A G and Gavriljuk V G 1999 {\it Acta
Mater.} {\bf 47} 927

\endrefs

\end{document}